\documentstyle[amssymb,aps,pre,multicol,epsf,psfig]{revtex}

\begin{document}
\draft
\author{M. C. Marques$^\dag$, M. A. Santos$^\ddag$ and J. F. F. Mendes$^\star$}
\title{Static critical behavior in the inactive phase of the pair contact process}

\address{Departamento de F\'\i sica and Centro de F\'\i sica do Porto, 
Faculdade  de  Ci\^{e}ncias, Universidade do Porto, \\
Rua do Campo Alegre 687, 4169-007 Porto, Portugal}

%\date{\today}

\maketitle

\begin{abstract}
Steady state properties in the absorbing phase of the $1d$ pair contact process (PCP) model are investigated. It is shown that, in typical absorbing states (reached by the system's dynamic rules), the density of isolated particles, $\rho_{1}$, approaches a stationary value which depends on the annihilation probability ($p$); the deviation from its 'natural' value at criticality, $\rho_{1}^{nat}$, follows a power law: $\rho_{1}^{nat}-\rho_{1} \sim (p-p_c)^{\beta_1}$ for $p>p_c$. Monte Carlo simulations yield $\beta_1=0.81$. A cluster approximation is developed for this model, qualitatively confirming the numerical results and predicting $\beta_1=1$.
The singular behavior of the isolated particles density in the inactive phase is explained using a phenomenological approach.
\end{abstract}


\begin{multicols}{2}

\section{Introduction}

In the simplest models undergoing absorbing state phase transitions in the directed percolation (DP) universality class, like the contact process (CP), the stationary state of the system in the inactive phase is the state devoid of particles. Other models in the same class, like the pair contact process (PCP), have however a richer structure, associated with the existence of an infinite number of absorbing states \cite{marro,hayeAP}. In the inactive phase, the absorbing state the system evolves to depends on the initial conditions and the distance from the critical point, and so does the average density of isolated particles in the stationary state. In the case of PCP, the field responsible for the dynamics (the density of pairs of particles, $\rho_{2}$) is coupled to another field (the density of isolated particles, $\rho_{1}$). This background of isolated particles is responsible for the non-universality of some dynamic properties of the system at criticality \cite{jensen,mendes}. Recently the one-dimensional PCP with particle diffusion (known as PCPD or annihilation/fission model) has received a lot of attention and was at the center of some controversy \cite{howard,carlon,haye,odor,park}; nevertheless its critical behaviour is not yet fully clarified. Carlon {\it et al.} \cite{carlon} presented a pair approximation study of this model which however is not valid for the non-diffusive case. The expected behaviour of PCP is qualitatively different from the $D=0$ limit of PCPD: in the non-diffusive case it is crucial to distinguish between {\it particles} and {\it isolated particles}, since all activity stops as soon as the number of pairs vanishes, whereas isolated particles may produce a pair if particle diffusion is allowed.

In PCP, as the system approaches the critical point from the active phase, the nonordering field approaches a `natural' value ($\rho_1^{nat}$), and its behavior is described by the same power laws as those of the order parameter \cite{gang4}. 

In the present work we  investigate the behavior of this model in the inactive phase by looking at properties of its {\it natural absorbing states}, the absorbing states selected by the system's dynamics. As shown by both mean-field-like approximations and Monte Carlo (MC) simulations, the stationary density of isolated particles develops a power law singularity as it approaches the critical point.

\section{Phenomenological approach}

In the inactive phase, the concentration of pairs is known to decay
 exponentially: $\rho_{2}(t) \sim e^{-\alpha t}$, with $\alpha \sim (p-p_{c})
^{\nu_{||}}$. The exponent $\nu_{||}$ is associated with the temporal
correlation length and mean-field theory predicts $\nu_{||}=1$.

Now, if one is interested in studying the stationary single particle 
concentration as a function of $p (>p_{c})$, one can refer to the coupled
Langevin equations describing this dynamical process \cite{munoz}. At the
mean-field level, the time evolution of $\rho_{1}$ is given by

\begin{equation}
\frac{d\rho_{1}}{dt}=r_{1} \rho_{2}-w_{1} \rho_{1} \rho_{2} -u_{1} \rho_{2}
^{2}
\end{equation}   
obviously coupled to the evolution of $\rho_{2}$. With the change of variables $\overline{\rho_{1}}=\rho_{1}-\frac{r_{1}}{w_{1}}$, equation (1) reads

\begin{equation}
\frac{d\overline{\rho_{1}}}{dt}=-w_{1} \rho_{2} \overline{\rho_{1}}
-u_{1} \rho_{2}^{2}.
\end{equation}

In the following, we will assume $\rho_{2}(t)=\rho_{2}(0) e^{-\alpha t}$.
Then, equation (2) can be solved exactly:

\begin{equation}
\overline{\rho_{1}}(t)=- \frac{u_{1} \alpha}{w_{1}^{2}}
\left( \frac{w_{1} \rho_{2}(0)}{\alpha} e^{-\alpha t} +1 \right) +
C e^{\frac{w_{1} \rho_{2}(0)}{\alpha} e^{-\alpha t}} 
\end{equation}

 with
\begin{equation}
C=\left( \overline{\rho_{1}}(0)+\frac{u_{1}}{w_{1}} \rho_{2}(0)+
\frac{u_{1} \alpha}{w_{1}^{2}} \right) e^{-\frac{w_{1} \rho_{2}(0)}{\alpha}}.
\end{equation}

In the limit $t\rightarrow \infty$, one has
\begin{equation}
\overline{\rho_{1}}(t\rightarrow \infty) = -\frac{u_{1} \alpha}{w_{1}^{2}}
+C. 
\end{equation}

In the vicinity of the critical point, when $\alpha \rightarrow 0$, $C$ 
decreases more rapidly than $\alpha$, and one can therefore conclude that
\begin{equation}
\overline{\rho_{1}}(t\rightarrow \infty ) \simeq - \frac{u_{1} \alpha}
{w_{1}^{2}}.
\end{equation}

At the critical point, $\overline{\rho_{1}}(t \rightarrow \infty )=0$, i.e. 
\begin{equation}
\rho_{1} (t \rightarrow \infty ) = \frac{r_{1}}{w_{1}}
\end{equation}

$\rho_{1}^{nat}$ is then given by $r_1/w_1$, as already found in
\cite{gang4}.
But now we can see that $\rho_{1} (t \rightarrow \infty )$ varies with $p$, 
in the inactive phase: 
\begin{equation}
\rho_{1}(t \rightarrow \infty ) \simeq \rho_{1}^{nat}
- \frac{u_{1}}{w_{1}^{2}} \alpha.
\end{equation}

We must point out that the present argument relies on a mean-field approach
and on the assumption that 
$\rho_{2}(t)=\rho_{2}(0) e^{- \alpha t}$ (which may not be true at early times). A variation of $\rho_{1} (t \rightarrow \infty )$ with $p$ is then
plausible. However, the prediction that the critical exponent of the
quantity $\rho_{1}^{nat} -\rho_{1}$ should be equal to $\nu_{||}$ is only
valid within a mean-field approximation and is not expected to apply to the 
$1d$ PCP model.

\section{Cluster approximation}

Mean-field-like kinetic equations for the PCPD were obtained in \cite{carlon}. 
Whereas the single-site approximation is only appropriate to the high 
diffusion limit ($D \rightarrow 1$), the pair approximation gives a good
qualitative picture of the model for $D>0$, but does not show some important
characteristics of the PCP without diffusion. Indeed, according to the pair
approximation, the steady state single particle density vanishes at the
critical point ($p_c=0.2$ for the pair approximation with $D=0$); however, it
 is well established that, at the critical point, the single particle density approaches a nonzero value, $\rho_{1}^{nat}$ ($\rho_{1}^{nat} \simeq 0.242$, in a
sequential dynamical process).
Also, the pair approximation predicts a power-law temporal decay for the pair
concentration in the inactive phase, contrary to the exponential decay found
in the simulations.

In the coarse-grained Langevin description, the fields that are used to
characterize the system configurations are the local pair density and the
local density of isolated particles. If one uses
 two-site clusters, as in the pair approximation, configurations with 
isolated particles are not treated appropriately. Indeed, if one aims at building a better approximation,
 one has to go to three-site clusters and consider $P_{010}$, the probability of having
an isolated particle at the center of the cluster. The price to pay is that
of increasing complexity: the number of variables and equations increases and
these have to be solved numerically; Carlon {\it et al.} dealt with two variables
 and two equations, so they were able to obtain some results analytically.
We have used a $(3,2)$ - cluster approximation \cite{ben} whose technical details are given in the Appendix.

In Fig.1, we have plotted the stationary values of $P_{11}$ and
 $P_{10}$ as functions of $p$.
As shown, the concentration of pairs vanishes for $p \geq 0.128$, i.e. 
$p_c=0.128$ within the present approximation. The estimate obtained by MC 
simulations is $p_c=0.0771$; so, as expected, the present result is an
 improvement when compared to what was obtained by the pair approximation.
More interestingly, the concentration of particles at the critical point
($\rho_{1}=P_{10}$, since $P_{11}=0$) is nonzero; indeed $\rho_{1}=0.229$, 
a value not far from $\rho_{1}^{nat}=0.2418(2)$, as obtained by the simulations \cite{dickman}.
On the other hand, a linear decrease of $\rho_{1}$ for $p>p_c$ can also
be noticed.

\section{Numerical simulations}

In this work, we concentrate on the critical behaviour of $\rho_{1}$ in
the inactive phase. 
The simulations that we present were done on systems of size $L=5000$; to ensure a sufficient number of pairs we chose an initial particle concentration of 0.5. The numbers of particles and of pairs were recorded versus time up to a number of Monte Carlo steps, $t_{max}$, which ranged from $t_{max}=10^{6}$ for $p$ very close to the transition ($p=0.0775$) down to $t_{max}=10^{4}$
for $p>0.085$. These times  were chosen such that most of the samples (typically around $20000$) had already entered the absorbing state. For each $p$ we
 evaluated the average activity time $\tau_{av}$. To obtain the stationary value of $\rho_1$, we have averaged the final number of particles in those samples which entered the absorbing state at a time around $\tau_{av}$. The short lived samples were excluded in order to eliminate finite size effects. On the other hand, in systems of this size, and for $p$ close to $p_{c}$, the
concentration of pairs is still considerably high (typically around $0.1$) at very long times - thus the concentration of isolated particles is lower than in samples where the number of pairs is vanishingly small. For this reason, we have also ignored the contribution from these long-lived samples. We have chosen to consider samples with $0.75\tau_{av}<\tau<1.25\tau_{av}$, but have checked that the results are stable with respect to other choices in a reasonable range.

In Fig.$2$ we show the stationary single particle concentration as a
function of $p$. Fig.$3$ shows a log-log plot of $(\rho_{1}^{nat} - \rho_{1})$ vs $(p-p_{c})$, with $\rho_{1}^{nat}=0.2418$ and $p_{c}=0.0771$.
A linear fit of the data is clearly appropriate and leads to an exponent
 $\beta_{1}=0.81(3)$. This seems to be a new exponent, not simply related
 to the DP exponents that describe the critical behaviour of other 
quantities, in the active phase of the PCP model.

\section{Conclusion}
In this work, we show that the approach to the critical point in the in inactive phase of the $1d$ PCP model 
is signaled by a power-law singularity of a static quantity - the deviation of the particle concentration from its 
critical point value, $\rho_1^{nat}$. A cluster-mean-field approximation was developed that predicts a finite value 
of $\rho_{1}^{nat}$, 
in reasonable agreement with MC simulations and a linear decrease of $\rho_{1}$ with $(p-p_c)$; this is confirmed by a 
phenomenological approach. 
Such a decrease in $\rho_{1}$ is also shown in our MC simulations; the best fit to the data is consistent with an 
exponent $\beta_1=0.81(3)$, different from the mean-field prediction.

Critical behavior of a static quantity in the inactive phase has been observed by Lipowski and Droz \cite{lipowski} in a rather different model with infinite number of absorbing states. In that case, a random walk argument leads to a relationship between the corresponding exponent and the order parameter exponent $\beta$. In the case of PCP such an argument cannot be applied and whether $\beta_1$ is related to the (DP) critical exponents that characterize the active phase of PCP requires further investigation. The same applies to other systems with infinitely many absorbing states. 

\acknowledgments
We thank  Ronald Dickman for a stimulating discussion and a critical 
reading of the manuscript and Adam Lipowski for communicating his 
work prior to publication. Partial support from project POCTI/1999/Fis/33141 
is acknowledged.

\section*{Appendix}

In the $(3,2)$-cluster approximation \cite{ben} one
uses $3-$site clusters and allows for an overlap of $2$ sites between adjacent
clusters. Thus, the probability of a $6-$site cluster in the state $ABCDEF$ 
is written as

\begin{equation}
P_{ABCDEF} = P_{ABC} \frac{P_{BCD}}{P_{BC}} \frac{P_{CDE}}{P_{CD}} 
\frac{P_{DEF}}{P_{DE}}.
\end{equation}

We use $x\equiv P_{111}$, $y\equiv P_{110}(=P_{011})$, $z\equiv P_{100}(=P_{001})$, $v\equiv P_{101}$ as independent variables. It can be easily shown that 
$P_{010}=v+z-y$, and $P_{000}=1-x-y-2v-3z$. We then study all the processes that can occur when a pair belonging to a
$6-$site cluster is selected. 

Take for example the configuration 
$1 \  1 \  1 \  1 \  1 \  1$. According to the above approximation, the
probability of this configuration is $A_{1}=x^4/c^3$, where $c=P_{11}
=x+y$.  
When the central pair is selected, then, with rate $p$, the configuration
$1 \  1 \  0 \  0 \  1 \  1$ is generated.
The variation in the number of $3-$site clusters in the state $1 \  1 \  1$ 
is then $\triangle x = -4$; analogously, the changes in the number of
$3-$site clusters in configurations $1 \  1 \  0 \  , 1 \  0 \  0$ and
$1 \  0 \  1$ are, respectively, $\triangle y = +1 , \triangle z = +1 ,
\triangle v = 0$.

On the other hand, according to the PCP dynamic rules, a pair can also
 create a new particle at a randomly chosen nearest neighbour provided this
is vacant.  If one considers the configuration 
 $1 \  \underline{1 \  1} \  0 \  1 \  1$ (whose probability is given, within
the present approximation, by $A_{11}=\frac{x y^{2} v}{c d^{2}}$, with $d\equiv P_{10}=
v+z$), then, with rate $1-p$, the configuration $1 \  1 \  1 \  1 \  1 \  1$
is generated. This change corresponds to $\triangle x = +3, \triangle y= -1 ,
\triangle z = 0 , \triangle v = -1$ .

The kinetic equations for $x,y,z,v$ are obtained by considering all the 
contributions from all the possible $6-$site configurations:
\begin{eqnarray}
\frac {dx}{dt}  & = & p [-4A_{1}-6A_{2}-4A_{3}-2A_{4}-4A_{6}-2A_{7}-2A_{9}]\nonumber\\
&+&(1-p) [3A_{11}+A_{12}+3A_{13}+2A_{14}+A_{15}+A_{16}\nonumber\\
&+& 2A_{17}+A_{18}] \nonumber\\
\frac {dy}{dt} & = & p [ A_{1}-A_{4}-A_{5}-2A_{7}-2A_{8}-2A_{9}-A_{10}]\nonumber\\
&+&(1-p) [-A_{11}-A_{13}]\nonumber
\end{eqnarray}

\begin{eqnarray}
\frac {dz}{dt} &=& p [A_{1}+2A_{2}+2A_{3}+A_{4}+A_{5}+2A_{7}-A_{10}]\nonumber\\
&+&(1-p) [-A_{12}-A_{15}]\nonumber\\
\frac {dv}{dt} &= &p [-2A_{3}-2A_{5}-2A_{7}-2A_{8}]\nonumber\\
&+&(1-p) [-A_{11}+A_{12}-A_{13}-A_{14}+A_{15}-A_{17}]\nonumber
\end{eqnarray}

where

\begin{displaymath}
A_{1}=\frac{x^{4}}{c^{3}}; \,\,\, A_{2}=\frac{x^{3} y}{c^{3}}; \,\,\, A_{3}=\frac{x^{2}vy}{c^{2}d}; \,\,\, A_{4}=\frac{x^{2} y^{2}}{c^{3}};
\end{displaymath}

\begin{displaymath}
A_{5}=\frac{y^2 v^2}{c d^2}; \,\,\,
A_{6}=\frac{x^2 yz}{c^2 d}; \,\,\,
A_{7}=\frac{x y^2 v}{c^2 d};  \,\,\,
A_{8}=\frac{y^{2} zv}{c d^{2}};
\end{displaymath}

\begin{displaymath}
A_{9}=\frac{x y^{2} z}{c^{2} d}; \,\,\,
A_{10}=\frac{y^{2} z^{2}}{c d^{2}}; \,\,\,
A_{11}=\frac{x y^{2} v}{c d^{2}};
\end{displaymath}

\begin{displaymath}
A_{12}=\frac{x y z^2}{c d (1-c-2d)};\,\,\,
A_{13}=\frac{y^3 v}{c d^2}; 
\end{displaymath}

\begin{displaymath}
A_{14}=\frac{x y v (v+z-y)}{c d^2}; \,\,\,
A_{15}=\frac{y^{2} z^{2}}{c d (1-c-2d)}; \,\,\,
\end{displaymath}

\begin{displaymath}
A_{16}=\frac{x y z (1-c-2d-z)}{c d (1-c-2d)}; \,\,\,
A_{17}=\frac{y^2 v (v+z-y)}{c d^2};
\end{displaymath}

\begin{displaymath}
A_{18}=\frac{y^2 z (1-c-2d-z)}{c d (1-c-2d)}.
\end{displaymath}

These equations were solved numerically.
\\
\\
$^\dag$ mcmarq@fc.up.pt\\
$^\ddag$ mpsantos@fc.up.pt\\
$^\star$ jfmendes@fc.up.pt
\\
\\

\end{multicols}

\newpage

\begin{figure}
\epsfxsize=80mm
\epsffile{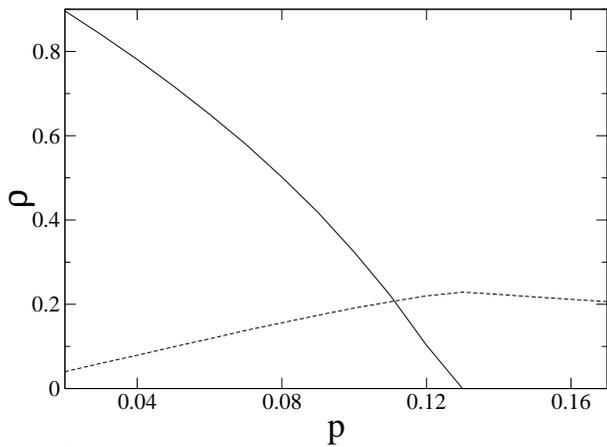}
\caption{ Pair density, $\rho_{2}$ ( \hrulefill\ ) and density of isolated particles, $\rho_{1}$ ($- - -$), as a function of $p$, within the $(3,2)$-cluster approximation.}
\label{fig1}
\end{figure}
\newpage

\begin{figure}
\epsfxsize=80mm
\epsffile{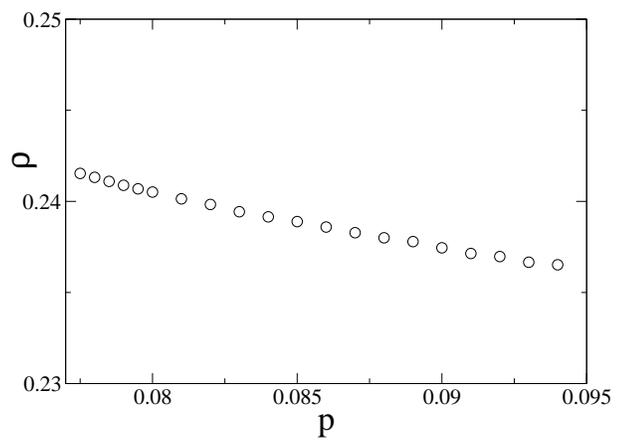}
\caption{Stationary particle density (in the inactive phase), as a function of $p$.}
\label{fig2}
\end{figure}

\newpage
\begin{figure}
\epsfxsize=80mm
\epsffile{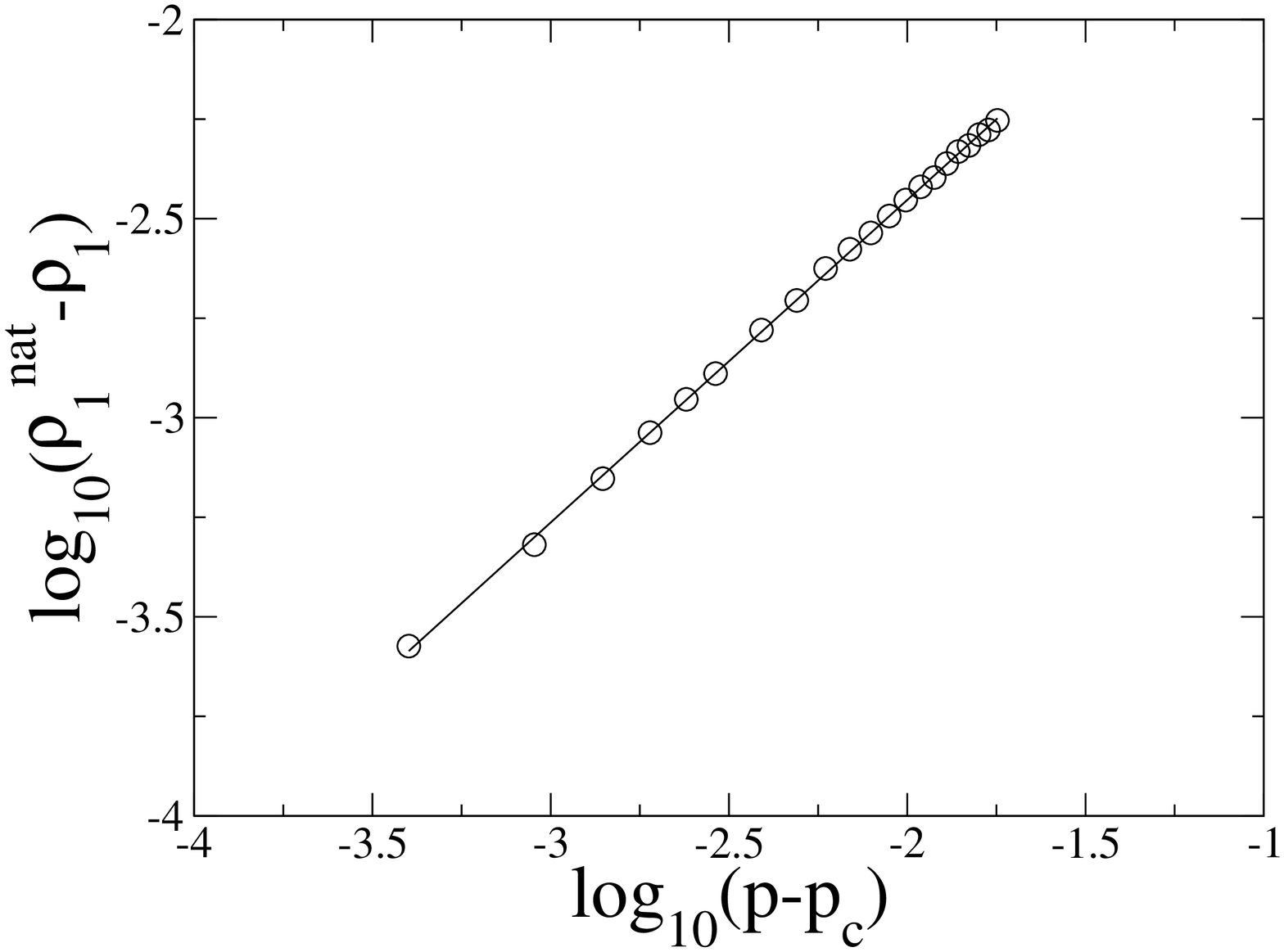}
\caption{Log-log plot of $(\rho_{1}^{nat} - \rho_{1})$ vs $(p-p_{c})$, with $\rho_{1}^{nat}=0.2418$ and $p_{c}=0.0771$. The straight line is a least squares
linear fit, with slope$=0.81$.}
\label{fig3}
\end{figure}

\end{document}